\documentclass[
 reprint,
superscriptaddress,
showpacs,preprintnumbers,
 amsmath,amssymb,
 aps,
 prl,
 lengthcheck,%
]{revtex4-1}

\usepackage{graphicx}
\usepackage{dcolumn}
\usepackage{bm}
\usepackage{hyperref}

\begin{document}


\title{Quantum oscillations and optical conductivity in Rashba spin-splitting BiTeI}

\author{C. Martin}
\affiliation{Department of Physics, University of Florida, Gainesville, Florida 32611, USA}
\author{E. D. Mun}
\affiliation{National High Magnetic Field Laboratory,
Los Alamos National Laboratory, Los Alamos, New Mexico 87545, USA}
\author{H. Berger}
\affiliation{Institute of Physics of Complex Matter, Ecole Polytechnique Federal de Lausanne, CH-1015 Lausanne, Switzerland}
\author{V. S. Zapf}
\affiliation{National High Magnetic Field Laboratory,
Los Alamos National Laboratory, Los Alamos, New Mexico 87545, USA}
\author{D. B. Tanner}
\affiliation{Department of Physics, University of Florida, Gainesville, Florida 32611, USA}

\date{\today}

\begin{abstract}
We report the observation of Shubnikov-de Haas (SdH) oscillations in single crystals of the Rashba spin-splitting compound BiTeI, from both longitudinal ($R_{xx}(B)$) and Hall ($R_{xy}(B)$) magnetoresistance. Under magnetic field up to 65 T, we 
resolved unambiguously only one frequency $F = 284.3\pm 1.3$ T, corresponding to a Fermi momentum $k_{F} = 0.093\pm 0.002$~\AA$^{-1}$.~The amplitude of oscillations is strongly suppressed by tilting magnetic field, suggesting a highly two-dimensional Fermi surface. Combining with optical spectroscopy, we show that quantum oscillations may be consistent with a bulk conduction band having a Rashba splitting momentum $k_{R}=0.046\pm$~\AA$^{-1}$.
\end{abstract}
\maketitle

\begin{figure}[htp]
\includegraphics[width=0.4\textwidth]{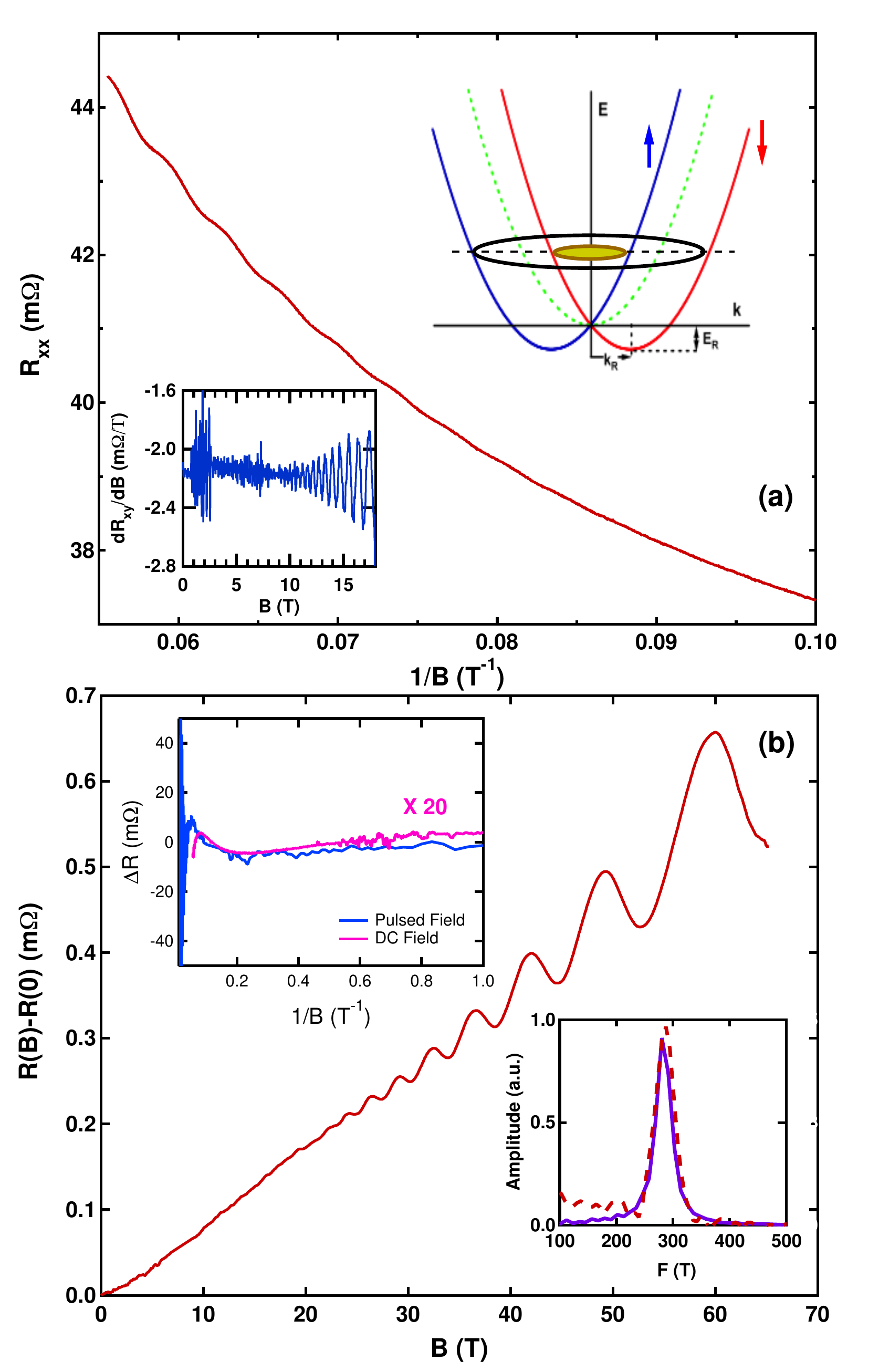}                
 \caption{(Color online) (a) Main panel: Longitudinal resistance $R_{xx}$~vs.~$1/B$ at T = 0.3 K for the magnetic field applied normal to sample surface. Upper inset: 1D representation of a Rashba spin-split conduction band showing the momentum ($k_{R}$) and the energy ($E_{R}$) splits. Lower inset: The field derivative of $R_{xy} (B)$, showing the presence of SdH oscillations at high magnetic field. The large noise between 1 and 3 T is due to magnetic flux jumps in the superconducting magnet. (b) Main panel: Magneto-resistance of BiTeI measured in pulsed magnetic field applied perpendicular to sample surface, at T = 4 K. Upper inset: Magneto-resistance obtained from two different measurements, in pulsed and DC magnetic field (multiplied by a factor of 20 for clarity), after subtracting a linear and quadratic background. Lower inset: FFT of SdH oscillations from pulsed (dashed line) and DC magnetic field (continuous line), respectively, scaled in amplitude for clarity.}  
\label{Fig:Data_Display}
\end{figure}

The claim of a large Rashba spin-splitting of the bulk electronic bands in BiTeI is based on a combination of theoretical calculations and photoemission spectroscopy (ARPES)~\cite{Ishizaka11}. Strong spin orbit interaction, originating from the presence of Bi with its large atomic number, and the absence of a center of inversion in the crystal structure give rise to a significant Rashba term in the Hamiltonian~\cite{Rashba60}, $H_{R}=\alpha_{R}\left(\hat{e}_{z}\times\vec{k}\right)\cdot\vec{S}$, where $\alpha_{R}$ is the Rashba parameter characterizing the strength of the effect, $\hat{e}_{z}$ is the direction along which the inversion symmetry is broken, $\vec{k}$ represents the momentum, and $\vec{S}$ is the spin of the electrons. The significance of the $\alpha_{R}$ parameter becomes more clear if we look at the effect of the Rashba term on the energy of a free electron system, which becomes $E_{\pm}=\hbar^{2}k^{2}/(2m^{*})\pm|\alpha_{R}|k$. The result is that electron energies are split between those with spin up (+) and spin down (-) in a plane perpendicular to $\hat{e}_{z}$, as sketched in the upper inset of Fig.~\ref{Fig:Data_Display}(a). The momentum and energy splitting both depend on the parameter $\alpha_{R}$.

The Rashba effect is of particular interest for the field of spintronics, where one aims to manipulate the spin of electrons for potential applications; moreover, a large value of $\alpha_{R}$ is very desirable. Values of $\alpha_{R}\approx 3$~eV\AA\ were found for asymmetric Bi/Ag(111) interfaces~\cite{Koroteev04, Ast07}. Recently, Ref.~\onlinecite{Ishizaka11} reported an even larger Rashba splitting, $\alpha_{R}=3.8$~eV\AA, in the bulk electronic bands of BiTeI. Optical spectroscopy of this compound found indeed an electronic excitation spectrum consistent with the splitting of the bulk conduction and valence bands~\cite{Lee11} and further photoemission study suggested the 3D nature of these bands~\cite{Sakano12}. More recent ARPES reports however, indicated the reconstruction of the band structure at the Te (or I) terminated surface and the existence of surface electronic branches, possibly with even larger Rashba spin-splitting~\cite{Landolt12,Crepaldi12}. On the theoretical side,~\textit{ab-initio} calculations for BiTeX(X=Cl, Br, I) do claim the formation of a surface 2D electron system distinct from the bulk states that has a larger Rashba splitting~\cite{Eremeev12}. 
 
Given that the fate of the surface states in BiTeI is still a debated issue and noting the particular sensitivity of photoemission experiments to the surface, we measured the in-plane longitudinal magnetoresistance $R_{xx}(B)$ and transverse (Hall) resistance $R_{xy}(B)$ in single crystals of BiTeI, searching for Shubnikov-de Haas oscillations as an alternative route to investigate the Fermi surface. Furthermore, we combine the results with optical reflectance data to understand better the origin of these oscillations.

Single crystals of BiTeI were grown by chemical vapor transport and Bridgman method. Two samples were initially screened and both revealed very similar quantum oscillations. Then, a complete study was performed on one sample with approximate dimensions 4$\times$6$\times$0.09 mm$^{3}$. Gold wires were attached using silver paint and the sample  resistance was measured using a commercial resistance bridge. The experiment was perfor.med in the SCM-2 facility at the National High Magnetic Field (NHMFL) in Tallahassee. The facility consists of a top loading $^{3}$He cryostat, with sample in liquid and a base temperature of 0.3 K, and an 18--20 Tesla superconducting magnet. Samples were mounted on a rotating probe with an angular resolution better than 1$^{\circ}$. Further magnetoresistance measurements up to 65 Tesla were also performed at the pulsed magnetic field facility of the NHMFL, in Los Alamos. Optical reflectance measurements were performed at the University of Florida. The data for frequencies between 30 and 5000 cm$^{-1}$ (4--620 meV), at temperatures as low as 10 K, were obtained using a helium flow cryostat mounted on a Bruker 113v Fourier spectrometer. Higher frequency reflectance, up to $\omega\approx 30000$~cm$^{-1}$ was measured at room temperature with a Zeiss microscope photometer and used to extrapolate the 10~K data for Kramers-Kronig analysis.

The main panel of Fig.~\ref{Fig:Data_Display}(a) shows the high magnetic field data for $R_{xx}$ at 0.3 K plotted against inverse field, obtained from the measurements in DC magnetic field up to 18 T. A small modulation, periodic in $1/B$ can be directly observed in the figure. Oscillations are also resolved in the transverse resistance $R_{xy}$, as can be seen from the lower inset of Fig.~\ref{Fig:Data_Display}(a), where $dR_{xy}/dB(B)$ is plotted. However, as shown in the main panel of Fig.~\ref{Fig:Data_Display}(b), the oscillatory behavior emerges undoubtedly in the magneto-resistance data above 25 T and their amplitude increases with magnetic field up to 65 T. Fourier transform (FFT) of these data yields a single frequency of oscillations, with the value $F = 284.3 \pm 1.3$~Tesla, for the field applied normal to the sample surface. In the lower inset of Fig.~\ref{Fig:Data_Display}(b), we compare the FFT frequencies obtained from DC and pulsed magnet filed, respectively; nearly perfect agreement can be observed.

This oscillation frequency is directly proportional to the area of the Fermi surface $S_{F}=2\pi eF/\hbar$ and furthermore to the Fermi momentum, for which we obtain $k_{F}=\left(S_{F}/\pi\right)^{1/2} = 0.093 \pm 0.002$~\AA$^{-1}$. Notably, this value of $k_{F}$ is comparable to that from some of the outer Fermi surfaces observed in photoemission experiments; it is nearly identical to the results from Ref.~\onlinecite{Ishizaka11, Sakano12}, which assign it to a bulk conduction branch and agrees within 50\% with the values from Ref.~\onlinecite{Crepaldi12}, but for electronic bands assigned to the surface. Moreover, it also agrees within better than 50\% with the value of $k_{F}$ for the surface states near the bottom of the conduction band, obtained from band structure calculations~\cite{Eremeev12}. 

Most importantly, this value of the Fermi momentum allows us to make an important observation about the magnitude of the Rashba splitting in BiTeI. When the chemical potential is situated above the Dirac cone, Rashba splitting of a conduction band should give rise to two Fermi surfaces, one associated with the outer and another with the inner branches, respectively, as is sketched in the upper inset of Fig.~\ref{Fig:Data_Display}(a). Therefore, two oscillation frequencies may be observed in SdH effect. On the other hand, if the chemical potential lies below the Dirac cone, one would expect a single, double degenerate frequency, corresponding to a momentum $k_{F}\leq k_{R}$, where $k_{R}$ is the Rashba momentum as shown in the upper inset of Fig.~\ref{Fig:Data_Display}(a). Most of the previously cited ARPES and theoretical studies seem to agree that in BiTeI, $k_{R}\approx 0.05$~\AA$^{-1}$. This value is about half of that obtained in our SdH study, clearly indicating that in our sample, the Fermi momentum is not situated below, but rather at the Dirac cone or slightly above, where the momentum associated with the outer Fermi surface becomes $k_{F}\geq 2\times k_{R}$. To further narrow the position of the chemical potential, we searched for possible hint of a low frequency oscillation, originating from the inner branch. In the upper inset of Fig.~\ref{Fig:Data_Display}(b) we show the sample magneto-resistance obtained, both in DC and in pulsed magnetic field, after subtracting a continuous background consisting of a linear and a quadratic term. There is indeed a hint of another modulation periodic with $1/B$, but with such a large period (low frequency), that only about half of a period can be resolved, even in a magnetic field as high as 65 T. It is also possible that the carrier density of this branch is so low that the quantum limit is reached at very low magnetic field. If we assign the peak and the dip marked with arrows in Fig.~\ref{Fig:Data_Display}(b) to half of a period, than the frequency would be no larger than 3T, which in turn, would place the Fermi energy in our sample less than 2 meV above the Dirac cone. Therefore, given that SdH effect is a robust measurement, we believe that we can estimate within about $1\%$ the Rashba momentum as $k_{R}=0.046\pm 0.0005$~\AA$^{-1}$. 

The angular dependence of the SdH oscillations provides a valuable insight into the dimensionality of the Fermi surface. For a 3D Fermi surface, electron orbits will be closed for any orientation of the magnetic field, and thus oscillations should, in principle, be observed for any angle between the magnetic field and sample surface. A quasi-2D Fermi surface (e.g., a cylinder) would show oscillations up to relatively large angles, provided that their frequency can be measured with the highest available magnetic field. In contrast, a strictly 2D layer backed by a conducting bulk may loose orbital coherence at small angles if a tilted field drives carriers into the bulk.
\begin{figure}[htp]
\includegraphics[width=0.4\textwidth]{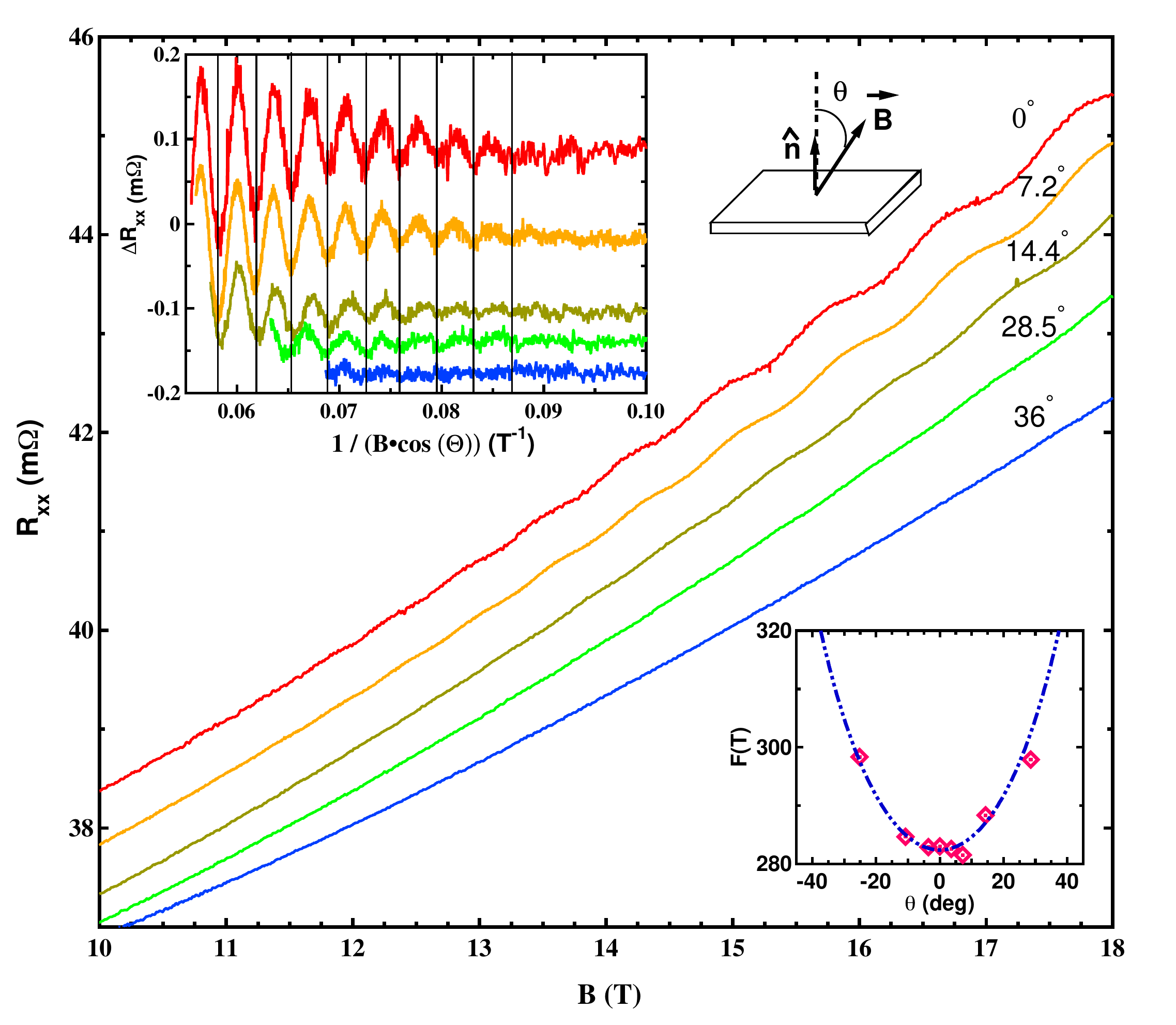}                
 \caption{(Color online) Main panel: $R_{xx} (B)$ above 10~T for different angles between the magnetic field and the sample surface. Upper inset: $\Delta R_{xx}$~vs.~the magnetic field component along the normal to sample surface and a sketch of the sample in tilted magnetic field. Lower inset: Angle dependence of the oscillation frequency (symbols) and fit to $1/\cos(\theta)$ (dashed line).}   
\label{Fig:Angle}
\end{figure}

The main panel of Fig.~\ref{Fig:Angle} shows the high magnetic field behavior (above 10 T) of $R_{xx}$ for different angles $\theta$ between the field and the sample surface at T = 0.3K, as sketched in the upper right inset. It can be visually observed that the oscillations are rapidly suppressed with increasing $\theta$ and that they are absent above $\theta\approx 30^{\circ}$. The angular dependence of their frequency, shown in the lower inset of Fig.~\ref{Fig:Angle}, has a $1/\cos(\theta)$ behavior, indicating a 2D character of the Fermi surface, whether it corresponds to bulk or surface electrons. In the upper inset of Fig.~\ref{Fig:Angle}, we display $\Delta R_{xx}$, obtained after background subtraction. First, we notice again the rapid decrease of oscillation amplitude with the angle of the magnetic field. Although not shown here, we measured up to $\theta > 90^{\circ}$, rotating the sample both directions with respect to the magnetic field and we confirm that oscillations only exists for $|\theta|\leq 30^{\circ}$. Second, we see that their period scales remarkably well with the component of magnetic field perpendicular to the sample surface ($1/B\cdot\cos(\theta)$). These behavior of SdH oscillations with angle is similar to that obtained from surface carriers of 2D structures~\cite{Cawiglia10} and of some of the topological insulators~\cite{Analytis10}, supporting the possibility of surface charge accumulation in BiTeI. On the other hand, BiTeI is a layered compound and a two-dimensional behavior may be induced, particularly at such low temperature, by stacking faults, similar to the observations in InSe compounds~\cite{Nicholas81}. Further studies of $c$-axis transport may shed light on this issue. Nevertheless, we will show later that optical reflectance, which is dominated by bulk properties, is in good agreement with the assumption that our observed oscillations originates from the bulk. Thus, from the SdH frequency, we calculate a 3D carrier concentration as $n_{3D}=\left(1/3/\pi^2\right)\left(2eF/\hbar\right)^{3/2}$ and obtain $n_{3D}=2.7\times 10^{19}$~cm$^{-3}$.

\begin{figure}[htp]
\includegraphics[width=0.45\textwidth]{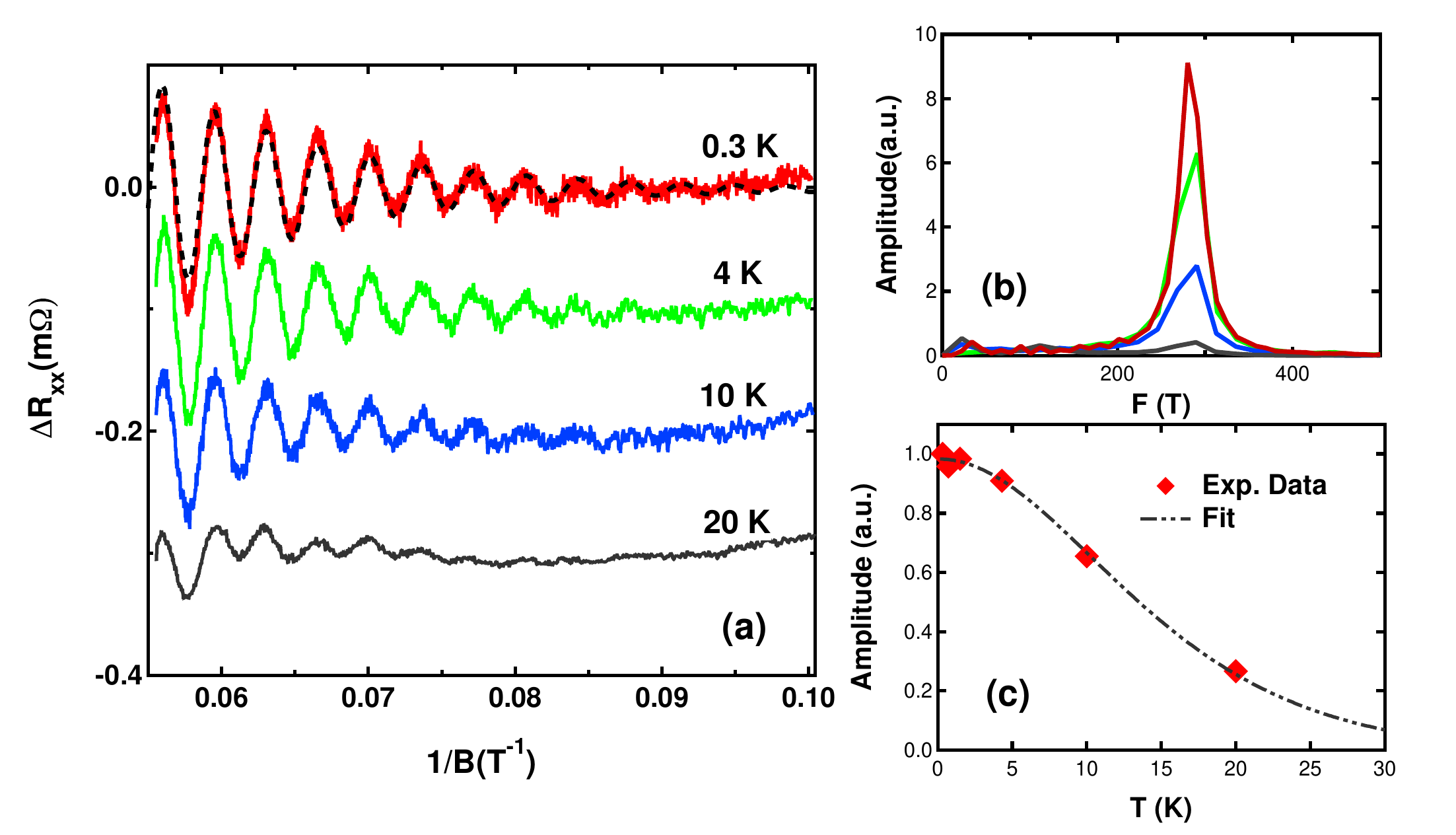}                
 \caption{(Color online) (a) $\Delta R_{xx}$ vs.$1/B$ (with $B$ above 10~T and applied perpendicular to the sample surface) at temperatures from 0.3 to 20~K. . The dashed line is a fit of the data at 0.3~K to the expression: $\exp(-\gamma T_{D})\cos(2\pi F/B+\pi)$ as explained in the text. (b) Fourier transform of the data from panel (a). (c) Amplitude of the oscillations at $1/B=0.06$~T$^{-1}$ for different temperatures, normalized to the value at 0.3~K (symbols) and a fit (dot-dash line) to the temperature-dependent damping term $\gamma T/\sinh(\gamma T)$, as explained in the text.}   
\label{Fig:Temp}
\end{figure}

To investigate further the carrier properties, we measured the temperature and magnetic-field dependence of the quantum oscillations. Figure~\ref{Fig:Temp}(a) shows the resistance $\Delta R _{xx}$ (after background subtraction) versus inverse field at different temperatures. Oscillations are visible up to at least 20 K, although significantly damped due to the thermal broadening of the quantized Landau levels. The effect of temperature is also evident if we look at the amplitude of the Fourier transform shown in Fig.~\ref{Fig:Temp}(b). This quantity is expected to follow the Lifshitz-Kosevich temperature dependence, $\gamma T/\sinh(\gamma T)$, with $\gamma = 14.69 m^{*}/m_0 B$, where $B$ is the magnetic field, $m^{*}$ is the effective mass, and $m_{0}$ the rest mass of the electron~\cite{Shoenberg84}. Figure~\ref{Fig:Temp}(c) shows that result of the fit to the above expression for the amplitude at $1/B = 0.06~T^{-1}$. We repeated the analysis for different values of $1/B$ and for the SdH oscillations from the Hall resistance, and obtained $m^{*} = 0.19\pm 0.02 m_{0}$. 

At fixed temperature, the amplitude of the SdH oscillations is enhanced at increased field (decreased $1/B$) as $\Delta R_{xx}\propto\exp(-\gamma T_{D})\cos(2\pi F/B+\pi)$, where $\gamma$ is defined above and $T_{D}$ is the Dingle temperature, $T_{D} = \hbar/\left(2\pi\tau k_{B}\right)$, related to the lifetime $\tau$ of the electrons~\cite{Shoenberg84}. In Fig.~\ref{Fig:Temp}(a) we show the fit of the data at 0.3 K, where it can be seen that both the amplitude and the phase are well reproduced by only considering one frequency. We obtain a Dingle temperature $T_{D}= 32 \pm 6$~K, which in turn gives a lifetime $\tau = 3.9 \pm 0.6 \times 10^{-14}$ s, and an estimated mobility $\mu=e\tau/m^{*}= 360 \pm 60$~cm$^{2}$/(V$\cdot$s). 

\begin{figure}[htp]
\includegraphics[width=0.45\textwidth]{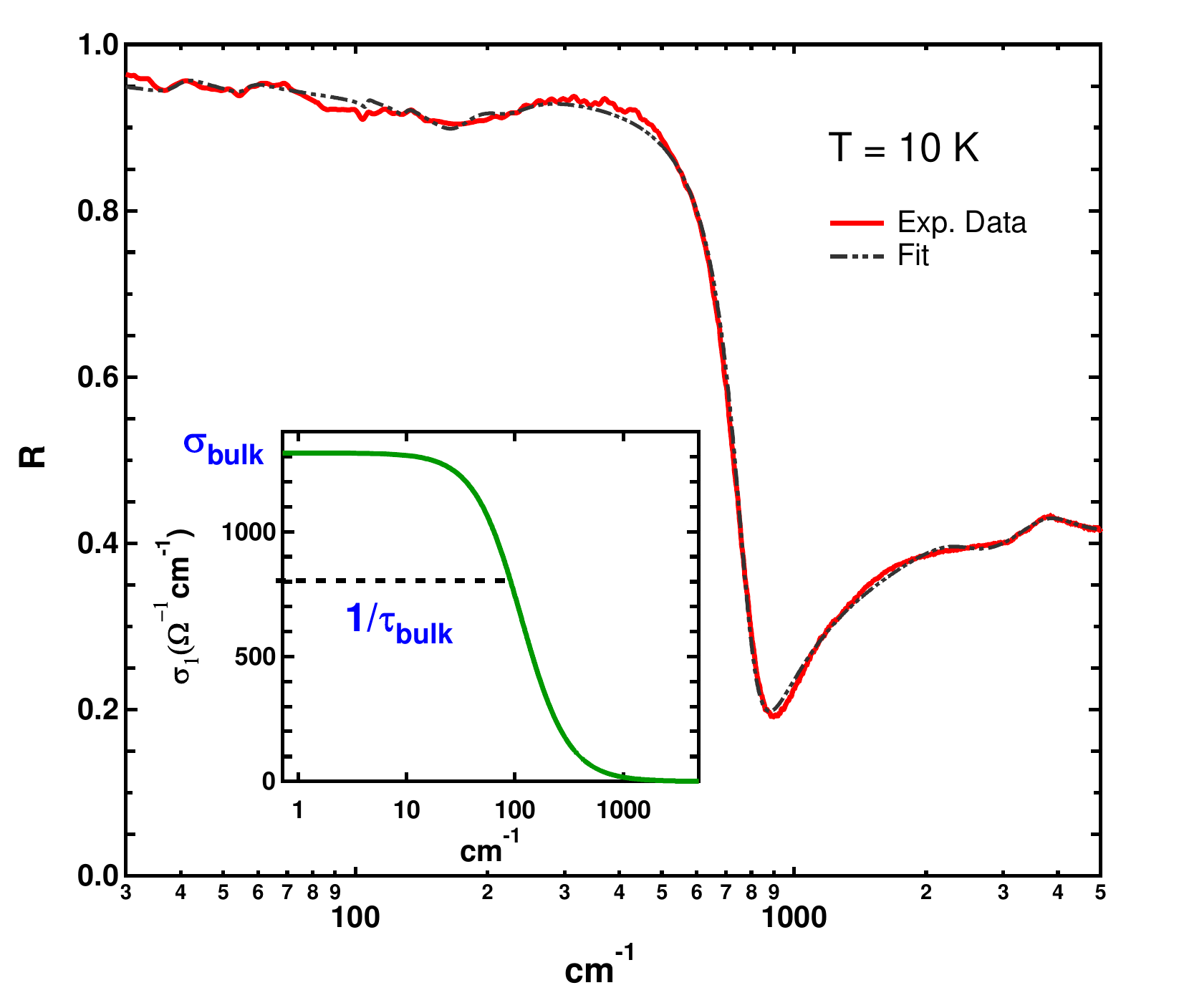}                
 \caption{(Color online) Main panel:~$\cal R(\omega)$ at $T = 10$~K (continuous red line) and a Drude-Lorentz fit (dashed dotted black line). Inset: Real part of optical conductivity obtained from the Lorentz-Drude fit, from which the bulk DC conductivity and the scattering rate of the bulk carriers are obtained.}   
\label{Fig:Refl}
\end{figure}

We investigated further the electronic properties of BiTeI by optical reflectance. The main panel of Fig.~\ref{Fig:Refl} shows the optical reflectance spectrum $\cal R(\omega)$ of the same sample at $T=10$~K. The data are similar to a previous optical study~\cite{Lee11}: there are several features at low frequency  associated with lattice vibrations, a clear sharp plasma edge at about 850 cm$^{-1}$ ($\approx 0.1$~eV) and broad structure at higher frequencies due to interband transitions. 
This reflectance is dominated by the bulk carriers. If we were to assume that the SdH oscillations originates from surface electrons, then they would create an impedance mismatch due to a surface impedance, which we estimated to $R_{\square} \approx 1$~k$\Omega$. A free-standing thin film with this impedance has a reflectance of
$\cal R$ = $(Z_{0}/R_{\square})^{2}/ (2+Z_{0}/R_{\square})^{2}\approx 0.02$, where $Z_{0} = 377$ $\Omega$ is the vacuum impedance~\cite{PalmerTinkham}. However, we see in Fig.~\ref{Fig:Refl} that at low frequencies $\cal R(\omega)$ $ \approx$ 0.96. Also, a few Angstrom-thick layer with a surface impedance of $R_{\square}~\approx 1$~k$\Omega$ would only attenuate the incident light by about 1\% in the frequency range of our measurements. Simulations show that for a conducting bulk (as we find) the addition of a monolayer-thick conducting surface layer changes the reflectance by less than 0.5\%. Therefore, most of the light probes the bulk. 

Analysis of optical reflectance~\cite{Wooten72} may use either Kramers-Kroning transformation or fits to a model such as the Drude-Lorentz model in order to estimate other optical quantities, such as the optical conductivity $\sigma(\omega)$. Here we have fit $\cal R(\omega)$ with the Drude-Lorentz model and the result is shown in Fig.~\ref{Fig:Refl} as the dash-dot line. The corresponding Drude contribution to the  conductivity $\sigma_{1}(\omega)=\sigma_{b}/(1+\omega^{2}\tau_{b}^{2})$ is shown in the inset. (The Kramers-Kronig-derived conductivity is very similar.) The lattice vibrations and interband transitions seen in Fig.~\ref{Fig:Refl} will be discussed in a separate work, here we focus on the free-carrier contribution to $\sigma_{1}(\omega)$. From the fit we obtain the scattering rate $1/\tau_{b}  = 118 \pm 5$~cm$^{-1}$ ($\tau_{b}\approx 4.5\times 10^{-14}$~s), the plasma frequency $\omega_{p}= 3030$~cm$^{-1}$ ($\approx~6\times 10^{14}$~rad/s) and, hence, the bulk conductivity $\sigma_{b} = 1300 \pm 80$~$\Omega^{-1}$cm$^{-1}$. 
The scattering rate for the bulk carriers, obtained from optical measurements agrees well with that obtained from Dingle temperature. Furthermore, if we assume an effective mass of $\approx 0.19 m_{0}$, determined above, then from plasma frequency $\omega_{p}^{2}=n_{b}e^{2}/(m^{*}_{b}\epsilon_{0})$, we calculate a bulk carrier concentration  
$n_{b}\approx 2\times 10^{19}$~cm$^{-3}$. This is also in fair agreement with the value from the temperature dependence of SdH oscillations, but it disagrees up to a factor of two with the result from Hall measurements. We believe however that the Hall measurements are more prone to errors, as the precise thickness where the current flows is difficult to know precisely, particularly in our relatively thick samples. Therefore, relying on quantum oscillations and optical reflectance, we may conclude that the bulk carrier concentration, corresponding to the chemical potential situated within 2 meV from the Dirac cone is $n_{b}= (2.35 \pm 0.35) \times 10^{19}$~cm$^{-3}$.

In conclusion, we measured Shubnikov-de Haas oscillations in the Rashba spin-splitting compound BiTeI. Under magnetic field as high as 65 T, we resolved unambiguously only one oscillation frequency and we showed that this is however consistent with a Rashba split conduction band, when the chemical potential is situated almost at the Dirac cone. We confirmed that the splitting momentum is $k_{R}=0.046\pm 0.0005$~\AA$^{-1}$. Although the quantum oscillations has a strongly two-dimensional character, optical reflectance, which probes mostly the bulk, suggests that their origin is likely from the bulk carriers.  

A portion of this work took place at the University of Florida,  supported by the DOE through Grant No. DE-FG02-02ER45984.  A portion was performed at the National High Magnetic Field Laboratory, which is supported by National Science Foundation Cooperative Agreement No. DMR-0654118, the State of Florida, and the U.S. Department of Energy. We would also like to thank Ju-Hyun Park, Glover Jones, and Timothy Murphy for support with the experiment at the National High Magnetic Field Laboratory. We also thank Dmitrii Maslov for stimulating discussions.

\end{document}